\documentclass[twocolumn,showpacs,preprintnumbers,amsmath,amssymb]{revtex4-1}
\usepackage{graphicx}
\usepackage{dcolumn}
\usepackage{bm}
\usepackage{color}
\begin{document}
%
%
\title{On the Landau theory of phase transitions}
%
%
%
%
%
\author{John Y. Fu}
%
%
\date{\today}
\affiliation{Department of Mechanical and Aerospace Engineering, The State University of New York, Buffalo, New York, 14260, USA}
%
%
%
\begin{abstract}
The Landau theory of phase transitions has been re-examined under the framework of a modified mean field theory in ferroelectrics. By doing so, one can see that there are two atomic movements involved in the ferroelectric phase transition; the first corresponds to the vibration of crystalline lattice, which will render phonon mode softening at the critical point, and the second represents the slow evolution of a partially ordered nematic phase formed by the cooperative behavior of high-temperature structure precursors. In this hierarchical dynamic structure, the former fast dynamics could be significantly modulated by the latter slow dynamics in the vicinity of the Curie temperature; it then turns out that it is the behavior of the nematic phase on approaching the critical point that makes the Landau theory deviate from experimental observations.
\end{abstract}
\maketitle
%
%
%
%
%
In 1937, Landau formulated an elegant theory, in which he proposed the concept of broken symmetry to explain continuous phase transitions; Landau also introduced a unique thermodynamic variable, i.e., the {\it order parameter}, to demonstrate his broken symmetry concept \cite{landau1937}. For the phase transition from paramagnetism to ferromagnetism, the spontaneous magnetization $M$ is defined as the order parameter. Thus, within the framework of the Landau theory, $M$ has the following values: $M=0$ if temperature $T$ is above the critical temperature $T_{c}$ and $M=\pm \ constant$ if $T$ is below $T_{c}$; the former ($T>T_{c}$) corresponds to one phase with higher symmetry and the latter ($T<T_{c}$) another phase with lower symmetry. In the absence of the external magnetic field and in the vicinity of $T_{c}$, the Landau free energy $F$ per unit volume can be written as follows \cite{landau1937}
\begin{equation}
F=F_{0}+\frac{1}{2}aM^{2}+\frac{1}{4}bM^{4}+\cdots,
\label{landau1937}
\end{equation}
where $F_{0}$ is the free energy that is independent of $M$; both $a$ and $b$ are coefficients; $a$ is also defined as $a=a_{0}t=a_{0}\frac{T-T_{c}}{T_{c}}$, in which $t=\frac{T-T_{c}}{T_{c}}$ is called the reduced temperature and $a_{0}$ is a coefficient. The Landau theory is one of the basic theoretical tools for describing phase transitions. By using the broken symmetry concept and the Landau free energy, continuous phase transitions can be explained qualitatively and quantitatively and the universal behavior of certain critical exponents near $T_{c}$ can be defined. For the ferromagnetic phase transition, the following critical exponents are given by the Landau theory:
\begin{eqnarray}
C & \propto & t^{-\alpha} \ \ (t>0) \ \ \Longrightarrow \ \ \alpha_{Landau}=0, \nonumber \\
C & \propto & (-t)^{-\alpha'} \ \ (t<0) \ \ \Longrightarrow \ \ \alpha'_{Landau}=0, \nonumber \\
\chi_{m} & \propto & t^{-\gamma} \ \ (t>0) \ \ \Longrightarrow \ \ \gamma_{Landau}=1, \nonumber \\
\chi_{m} & \propto & (-t)^{-\gamma'} \ \ (t<0) \ \ \Longrightarrow \ \ \gamma'_{Landau}=1, \nonumber \\
M & \propto & (-t)^{\beta} \ \ (t<0) \ \ \Longrightarrow \ \ \beta_{Landau}=\frac{1}{2}, \nonumber \\
H_{m} & \propto & M^{\delta} \ \ (t=0) \ \ \Longrightarrow \ \ \delta_{Landau}=3,
\label{criticalexponents1}
\end{eqnarray}
here $C$ and $\chi_{m}$ represent specific heat and magnetic susceptibility, respectively; $H_{m}$ represents the external magnetic field. Unfortunately, these estimated results are not in good agreement with experimental observations. The corresponding experimental data and the exponents given by the Landau theory are summarized in Table 1 for convenience.

The major criticism of the Landau theory can be concisely summarized as follows: thermal fluctuations are neglected in the Landau theory; however, near the critical temperature, thermal fluctuations become important and render the Landau theory inaccurate \cite{lubensky2000}. Before we discuss this common belief, it is perhaps worth briefly discussing the mean field approximation of the Landau theory so that one can see what else, in addition to thermal fluctuations, is missing in the Landau theory.

Let us consider a uniaxial magnetic material with the simple cubic structure. Its Hamiltonian can be written, via the Ising model, as
\begin{equation}
H=-J\sum_{\langle i,j\rangle}s_{i}s_{j}-h\sum_{i}s_{i},
\label{mft1}
\end{equation}
where $H$ and $s_{i}=\pm1$ represent the Hamiltonian and the Ising spin on the $i$th lattice site, respectively; $s_{j}$ is defined as the neighboring Ising spin of the $i$th site; $h$ is the external magnetic field on the $i$th site; $J$ represents the interaction strength between the nearest neighbors. If we assume that the mean value of spin, $m$, is site-independent, i.e., $m=\langle s_{i}\rangle=\langle s_{j}\rangle$, then $\delta s_{i}=s_{i}-m$ and $\delta s_{j}=s_{j}-m$ represent the random values caused by thermal fluctuations. Substitute both $s_{i}$ and $s_{j}$ by $s_{i}=m+\delta s_{i}$ and $s_{j}=m+\delta s_{j}$ in the above equation, the Hamiltonian can be simplified as follows \cite{lubensky2000}
\begin{equation}
H\approx H_{MF}=\frac{Jm^{2}Nz}{2}-\left(h+zJm\right)\sum_{i}s_{i},
\label{mft2}
\end{equation}
and the corresponding partition function $Z$ can also be obtained as
\begin{equation}
Z=e^{-\frac{Jm^{2}Nz}{2k_{B}T}}\left[2\mathrm{cosh}\left(\frac{h+zJm}{k_{B}T}\right)\right]^{N},
\label{mft3}
\end{equation}
where $N$ is the total number of lattice sites; $z$ is the number of nearest neighbors of each spin; $k_{B}$ represents the Boltzmann constant. Using Eqs. (\ref{mft2}) and (\ref{mft3}), one can further determine that $m=0$ when $T>T_{c}$ or $m=\pm \ constant$ when $T<T_{c}$ \cite{lubensky2000}. After such a simplification, one can see that the Hamiltonian of a many-body system given in Eq. (\ref{mft1}) can be reduced to that of an effective one-body system given in Eq. (\ref{mft2}), in which the second order fluctuation term, $\delta s_{i}\delta s_{j}$, is neglected and $\delta s_{i}$ and $\delta s_{j}$ are split and integrated into $h_{eff}=h+zJm$; here, $h_{eff}$ represents an effective mean field or molecular field on the $i$th site, which arises both from the external field, $h$, and from the exchange field induced by the neighboring spins, $zJm$. The expression of $h_{eff}$ can be further extended to the corresponding effective macroscopic mean field, $H_{eff}=H_{m}+\alpha M$; here $\alpha$ is a coefficient and $M$ represents the total of $m$ within the material. The above mean field approximation of the Ising model reflects a part of the nature of the Landau theory, which is based on an assumption that the considered material undergoes harmonic vibrations when perturbed by external stimuli (including thermal fluctuations). Under this situation, the atoms sitting on lattice sites or the corresponding Ising spins on lattice sites of the considered material can be regarded as distinguishable physical quantities that obey the Boltzmann distribution; the mutual interactions between the atoms or the spins due to thermal fluctuations can be regarded as random variables and their net effect can be neglected. Thus, the above-mentioned simplified Hamiltonian and the partition function can be derived. Within this mean field approximation, the Landau theory assumes that the net value of total spins is a site-independent non-zero constant value $M$ and the mutual interactions between spins are zero when $T<T_{c}$ or both of them are zero when $T>T_{c}$.

\begin{figure}[h!]
\begin{center}
\includegraphics[width=1.0\columnwidth]{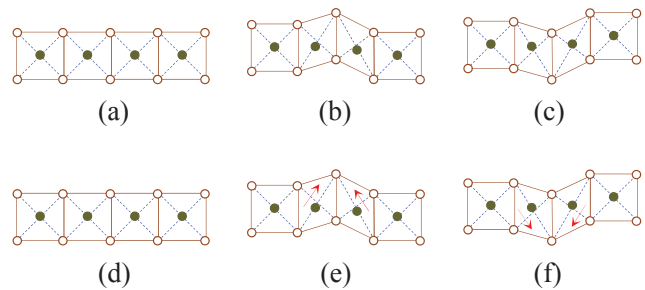}
\caption{Diagrammatic representation of the emergence of distorted crystal lattice; (a) the undeformed magnetic crystal lattice; (b) and (c) the distorted magnetic crystal lattice; (d) the undeformed crystal lattice of a perovskite material in the paraelectric state, in which there is no net dipole moment; (e) and (f) the distorted perovskite crystal lattice with net dipole moment represented by arrows.}
\label{distortedlattice}
\end{center}
\end{figure}

However, the lattice vibration of the considered material is by no means harmonic all the time, especially when approaching the critical point. If the anharmonic lattice vibration induced by thermal fluctuations makes the crystalline lattice distorted locally in the considered material, the assumption of the Landau theory is still valid. In this case, the spins on the distorted lattice sites might not obey the Boltzmann distribution but the net value of those spins and the mutual interactions between them can be assumed to be zero since those distorted lattice sites are randomly distributed in the spatial domain. Within the framework of the Landau theory, the corresponding Hamiltonian has a global symmetry since it is supposed to be invariant with respect to spatially uniform group operations. For those spins on the distorted lattice sites and the mutual interactions between them, their net contribution to the magnetization is zero and thus they do not affect the global symmetry of the Hamiltonian. If, for some reason, the spins and the interactions cannot be averaged out and do form a global magnetization in both $T<T_{c}$ and $T>T_{c}$ cases (of course, such a magnetization, if exists, must be weak), they can indeed affect the global symmetry and then further shift the phase transition behavior of the considered material. This situation has not been considered in the Landau theory. Furthermore, there is another concern in the Landau theory. In thermodynamics, material properties are described by using continuous functions so that the considered material must be treated as a continuum \cite{lubensky2000}. Since the Landau theory is a thermodynamic model, it must be subject to this thermodynamic limit; such an essential prerequisite is implicitly assumed to be only missing at $T_{c}$ but satisfied above and below $T_{c}$ in the Landau theory due to the assumption of the global symmetry. It later becomes clear that it is these two factors that render the Landau theory inaccurate and alter the asymptotic behavior of the corresponding physical quantities defined in Eqs. (\ref{criticalexponents1}).

\begin{figure}[h!]
\begin{center}
\includegraphics[width=1.0\columnwidth]{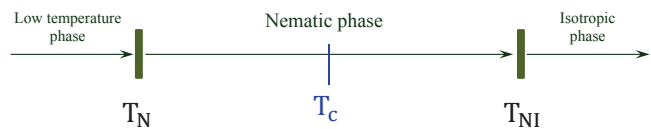}
\caption{Schematic representation of the cooperative behavior spectrum of HTSPs; below temperature $\mathrm{T_{N}}$, the quantity of disordered structures is too small to behave cooperatively; above temperature $\mathrm{T_{NI}}$, thermal energy is large enough to destroy any possible cooperative behavior of disordered structures; between $\mathrm{T_{N}}$ and $\mathrm{T_{NI}}$, a nematic phase as the ensemble of disordered structures exists.}
\label{nematicphase}
\end{center}
\end{figure}

Since the above-mentioned spins and interactions are related to the distorted crystal lattice, we need to take a close look at its formation and the possible dynamic behavior of the corresponding microscopic disordered structures under thermal fluctuations. For a crystalline or polycrystalline material, it has a purely ordered state at temperatures just above absolute zero and a completely disordered state above its melting point; at temperatures between those two extremes, the material should have both ordered and disordered states and the formation of the latter is largely due to thermal fluctuations (for simplicity, we here do not consider defects or microscopic disordered structures generated during material manufacturing processes). For the normal crystal lattice of magnetic materials shown in Fig. 1 (a), even at temperatures far below the melting point, there always exists the probability that certain atoms could gain extra kinetic energy from thermal fluctuations to move quasi-permanently away from their original equilibrium positions and make the crystal lattice distorted as shown in Fig. 1 (b) or (c). Such distorted lattice could change the local electron degeneracy and then makes the local spins and the effective local magnetic field very complicated in the considered magnetic material. In order to simplify the theoretical analysis to be discussed, we will mainly consider perovskite materials and the ferroelectric phase transition in the rest of the paper. For simplicity, we will first consider a perovskite material in the paraelectric state. For the undeformed crystal lattice of such a material shown in Fig. 1 (d), one can imagine that the centers of gravity of the negative and the positive ions coincide leading to zero net dipole moment; for the distorted lattice shown in Fig. 1 (e) or (f), however, there is relative displacement between the centers due to the local non-uniform deformation, which results in net dipole moment. The local disordered structures corresponding to such distorted crystal lattice are defined as high-temperature structure precursors (HTSPs). Since the formation of the distorted structures is caused by thermal fluctuations, their quantity is proportional to temperature. Therefore, as temperature rises, the quantity also increases. At a certain temperature $T_{N}$, the quantity of HTSPs has reached a threshold or the effective distance between HTSPs has been reduced below a critical value so that HTSPs will start to interact with each other, which can be regarded as their cooperative or self-organization behavior, and then form a unique nematic phase. The driving force behind this structural transformation is the competition between energy and entropy. Let us consider a crystalline perovskite material containing HTSPs; its Gibbs free energy in the absence of external fields can be written as $G=U-TS$. Clearly, the internal energy $U$ will increase due to the increment of the local strain and electric potential energy generated by HTSPs shown in Fig. 1 (e) or (f). If $T$ remains unchanged or changes slowly, the entropy $S$ must rise to reduce $G$. The simplest way to increase $S$ is that the chemical bonds between the atoms on the distorted lattice sites are partially broken so that the corresponding disordered structures could gain more freedom to rotate and are oriented along local preferred directions to form a nematic phase, which leads to a decrease in the orientational entropy but an increase in the positional entropy and, eventually, results in a net increase in the total entropy. In addition, since the nematic phase has no polarity (the cylindrical symmetry $\mathrm{D_{\infty h}}$) \cite{blinov2010}, the induced electric potential energy shown in Fig. 1 (e) or (f) will decrease when the phase is formed so that $G$ could be further reduced. If we use $v$ to represent the local preferred direction and approximately treat HTSPs as molecules, then their orientational order parameter, $S_{op}$, can be written as \cite{lubensky2000,blinov2010}
\begin{equation}
S_{op}=\frac{1}{2}\langle3\left(v^{i}, \vec{n}\right)^{2}-1\rangle=\frac{1}{2}\langle\left(3\mathrm{cos}^{2}\theta^{i}-1\right)\rangle,
\label{orderparameter}
\end{equation}
where $\langle\ \rangle$ represents the average; $v^{i}$ is defined as the given direction of the disordered structure located at the position $i$; $\vec{n}$ is usually called the director that represents a particular direction; $\theta^{i}$ is defined as the angle between $v^{i}$ and $\vec{n}$ at the position $i$. If $0<S_{op}<1$, we can say that HTSPs cooperatively form a nematic phase. $S_{op}=1$ corresponds to an ideal case, in which all HTSPs are perfectly aligned. If temperature continues to rise, at a certain point $T=T_{NI}$, the thermal energy will be large enough to disturb HTSPs, which makes HTSPs randomly oriented and then renders $S_{op}=0$. Therefore, when $T>T_{NI}$, all HTSPs behave like a normal liquid, which has an isotropic phase. The cooperative behavior spectrum of HTSPs is shown schematically in Fig. 2.

\begin{figure}[h!]
\begin{center}
\includegraphics[width=1.0\columnwidth]{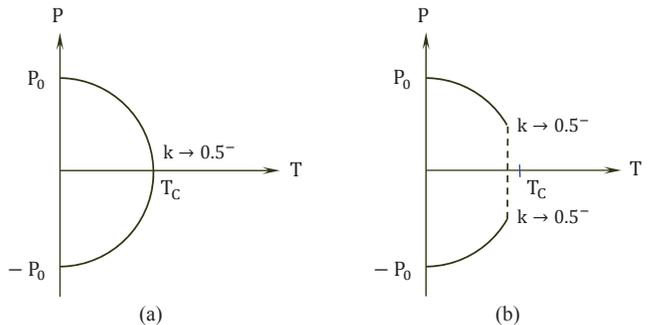}
\caption{Schematic representation of how the dynamic behavior of $k$ could alter phase transitions; (a) $k\rightarrow0.5^{-}$ coincides with $\mathrm{T_{C}}$ in an allowed temperature range, then the ferroelectricity-to-paraelectricity phase transition will occur at $\mathrm{T_{C}}$; (b) $k\rightarrow0.5^{-}$ occurs far before $\mathrm{T}$ reaches $\mathrm{T_{C}}$, then there exists no second order phase transitions.}
\label{orderparameter}
\end{center}
\end{figure}

We now consider the behavior of this nematic phase in a perovskite material in the ferroelectric state. Since HTSPs are usually metastable, the built-in polarization field $E_{b}$ (in the absence of the external electric field) would tend to disturb them. According to Le Chatelier's principle, the nematic phase would then undergo a specific structural change to counteract any imposed change by the filed \cite{landau1980}. Therefore, the effective built-in polarization field inside the specimen will be $E_{b}^{eff}=E_{b}-\tilde{\varrho}P_{h}=\varrho P-\tilde{\varrho}P_{h}$; here the negative sign is due to Le Chatelier's principle; both $\varrho$ and $\tilde{\varrho}$ are coefficients, respectively; both $P$ and $P_{h}$ represent the spontaneous polarization (the order parameter) and the induced counteractive polarization of the nematic phase, respectively. For the sake of convenience, we can further write $E_{b}^{eff}=\varrho(1-k)P$ and define $k$ as
\begin{equation}
k=\frac{k_{0}S_{op}(T-T_{N})}{T_{N}} \ \ \ \mbox{$(T_{N}<T<T_{NI})$},
\label{niphasetransition}
\end{equation}
where both $k$ and $k_{0}$ are dimensionless coefficients. For a unit-volume ferroelectric material containing HTSPs, in the absence of the external electric field, the corresponding free energy $F$ up to the fourth order of $P$ can be written as
\begin{equation}
F=F_{0}+\frac{1}{2}a(1-k)P^{2}-\frac{1}{3}\eta_{h}P_{h}^{3}+\frac{1}{4}b(1-k)^{2}P^{4},
\label{modifiedlandau0}
\end{equation}
$\eta_{h}$ is a coefficient. This equation represents a special case of the Landau-de Gennes free energy \cite{degennes1971}, which indirectly demonstrates the co-existence of the ferroelectric phase transition and a weak first order structural transformation. If $\eta_{h}$ is very small and can be neglected, then the above equation will reduce to a modified Landau free energy, which is written below
\begin{eqnarray}
F & = & F_{0}+\frac{1}{2}a(1-k)P^{2}+\frac{1}{4}b(1-k)^{2}P^{4} \nonumber \\
  & = & F_{0}+F_{L}+\left(-\frac{ak}{2}P^{2}+\frac{b}{4}k^{2}P^{4}-\frac{bk}{2}P^{4}\right),
\label{modifiedlandau}
\end{eqnarray}
where $F_{L}=\frac{a}{2}P^{2}+\frac{b}{4}P^{4}$. In the above equation, one can see that $k$ actually represents, in the statistical sense, the fraction of the electric potential energy, which is generated by the cooperative movement of HTSPs to reduce the potential energy brought by the order parameter. Now we can briefly summarize what we have done so far: the potential energy corresponding to both the crystalline phase and the nematic phase as the ensemble of HTSPs or, more precisely, the mutual interactions between the atoms of HTSPs is integrated together by using a modified mean field theory (MMFT), in which HTSPs are treated as a partially ordered liquid. In this way, two phases, the crystalline phase and the nematic phase, must be considered together since the latter is constrained by the former. The dynamics of the crystalline phase is fast, which corresponds to the vibration of crystalline lattice. On the other hand, the symmetry group in the nematic phase is the continuous rotation group, which has an uncountable continuum of symmetry elements that renders the corresponding relaxation extremely long and, thus, its dynamic behavior is extremely slow \cite{lubensky2000}. Therefore, the integration of these two phases does not violate Landau's global symmetry assumption in ferroelectric materials and his concept of broken symmetry in the ferroelectric phase transition; however, it does embed a weak first order structural transformation in the second order phase transition, which is partially reflected by the term $(1-k)$ in the modified Landau free energy given in Eq. (\ref{modifiedlandau}). Before discussing how this term could alter the phase transition behavior, we need to find out the meaning of $k$. For a crystalline ferroelectric material, $k=0$ when $T<T_{N}$, which corresponds to the situation that Eq. (\ref{modifiedlandau}) will reduce to the regular Landau free energy, and $k$ will increase when $T$ rises. When $T_{N}<T<T_{NI}$, the nematic phase of HTSPs will start to emerge in the considered material and the chemical bonds between certain atoms of HTSPs are partially broken so that they cannot be considered the integral part of the crystalline lattice of the considered material. Therefore, $k=0.5$ represents, in the statistical sense, the thermodynamic limit; below this limit ($k<0.5$), the material is still a continuum; above this limit ($k>0.5$), the material cannot be regarded as a continuum anymore \cite{johnyfu2012}. Thus, from the viewpoint of thermodynamics, Eq. (\ref{modifiedlandau}) is only valid when $k<0.5$. Such a situation has been reflected in the Coffin-Manson law, which is written below
\begin{equation}
\frac{\epsilon_{p}}{2}\approx\epsilon_{f}(2N_{f})^{-(1-k)}=\epsilon_{f}(2N_{f})^{-\beta_{CM}},
\label{coffinmanson3}
\end{equation}
where $\epsilon_{p}$ is the induced plastic strain; $\epsilon_{f}$ is defined as the ductility coefficient; $N_{f}$ represents the fatigue life, and $\beta_{CM}$, which has been proved to be $\beta_{CM}=1-k$ \cite{johnyfu2012}, is called the Coffin-Manson exponent. It has been observed that $\beta_{CM}$ possesses the remarkable universality; $\beta_{CM}\sim0.5$ has been found in single-phased metallic materials (see Ref. \cite{johnyfu2012} and the references cited therein). $\beta_{CM}>0.5$ or $k<0.5$ can be interpreted that the test specimen undergoes elastic deformation (or the specimen is still a continuum); whereas $\beta_{CM}<0.5$ or $k>0.5$ demonstrates that fracture and crack have emerged and started to grow in the specimen (in other words, the specimen is not a continuum anymore). Therefore, the Coffin-Manson law actually describes the fatigue behavior near the limit where fracture and crack will emerge in the specimen.

We now consider the situation, $k\rightarrow0.5$, in the ferroelectricity-to-paraelectricity phase transition. From the above discussion, it is clear that the crystalline phase and the nematic phase form a hierarchical structure. At very low temperatures, the atomic movement of the former is frozen, then the latter does not exist; as temperature rises or the movement of the former becomes fast, the latter starts to emerge. Therefore, the latter is constrained by the former. At the critical point $T_{c}$, the crystalline phase collapses due to phonon mode softening and the considered ferroelectric material must be in the disordered state. The nematic phase could quickly grow up if the temperature does not rise too fast. Thus, $k$ must reach the thermodynamic limit at some point near $T_{c}$ (without loss of generality, we assume $T_{N}<T_{c}<T_{NI}$ as shown in Fig. 2). This situation brings a rather subtle change to the phase transition. As shown in Fig. 3, if $k\rightarrow0.5^{-}$ coincides with $\mathrm{T_{C}}$ in a very small temperature range, then the ferroelectricity-to-paraelectricity phase transition will occur; if, however, $k\rightarrow0.5^{-}$ occurs far before $\mathrm{T}$ reaches $\mathrm{T_{C}^{-}}$, then there exists no second order phase transitions. We shall demonstrate how the dynamic behavior of $k$ would alter the phase transition and the corresponding critical exponents in the former case.

By taking the partial derivative of $F$ given in Eq. (\ref{modifiedlandau}) with respect to $P$ and then letting the result be equal to zero, we have the following equation
\begin{equation}
\frac{\partial F}{\partial P}=(1-k)P\left[a+b(1-k)P^{2}\right]=0.
\label{stable1}
\end{equation}
One of the solutions of this equation is given below:
\begin{equation}
P=\pm\left(-\frac{a_{0}}{b}\right)^{\frac{1}{2}}\left(-\frac{t}{1-k}\right)^{\frac{1}{2}}. \ \ \ \mbox{$t<0$}
\label{betaexponent1}
\end{equation}
Contrary to the Landau theory, the phase transition characterized by the free energy given in Eq. (\ref{modifiedlandau}) involves two atomic movements near $T_{c}$ as previously mentioned; the result of the fast dynamic behavior can be interpreted by the soft-mode theory \cite{cochran1960,cochran1961} but it could be significantly modulated by slowly fluctuating HTSPs in the nematic phase in the vicinity of $T_{c}$. Therefore, near $T_{c}$, the critical exponent $\beta$ is determined by both movements as shown in the above equation, in which both $t$ and $1-k$ have key influence on $\beta$. To estimate $\beta$, one has to do the following modification. We first rewrite $(1-k)P^{2}$ as $(1-k)P^{2}=1+(1+k)\left[\frac{1-k}{1+k}P^{2}-\frac{1}{1+k}\right]$; near $T_{c}$, $P\ll1$, $k\rightarrow1$ (mathematically, $k$ approaches 1 at $T_{c}$ but, physically, $k\rightarrow0.5$ when $\mathrm{T}$ is reaching $\mathrm{T_{C}}$; beyond this limit, the equations written here cannot be used), and $\frac{1-k}{1+k}<1$, thus $\left|\frac{1-k}{1+k}P^{2}-\frac{1}{1+k}\right|<1$. By taking advantage of the binomial series expansion, we have the following relationship: $(1-k)P^{2}\approx P_{m}^{2+2k}$; here $P_{m}^{2}=\frac{1-k}{1+k}P^{2}-\frac{1}{1+k}+1=\frac{1-k}{1+k}P^{2}+\frac{k}{1+k}$. Eq. (\ref{betaexponent1}) can then be rewritten as
\begin{equation}
P_{m}=\pm\left(-\frac{a_{0}}{b}\right)^{\frac{1}{2+2k}}\left(-t\right)^{\frac{1}{2+2k}}. \ \ \ \mbox{$t<0$} \\
\label{betaexponent2}
\end{equation}
The mathematical expression of $\beta$ is given by
\begin{equation}
\beta=\frac{1}{2+2k^{-}},
\label{beta}
\end{equation}
where $k^{-}\equiv k\rightarrow0.5^{-}$. If $k=0$, $\beta$ will become $\beta_{Landau}$. Thus, the upper limit of $\beta$ is $\frac{1}{2}$; since $k<0.5$, its lower limit should be $\frac{1}{3}$.

\begin{table*}
\caption{Critical exponents of continuous phase transitions; the values with the superscript symbol $[a]$ in the following table are taken from Table 3.1 of Ref. \cite{goldenfeld1992} and the references cited therein; the ones with $[b]$ are taken from Table 5.4.2 of Ref. \cite{lubensky2000} and the references cited therein; the ones with $[c]$ are the simulation results of the three-dimensional renormalization-group of Ising systems reported in Refs. \cite{zinnjustin1977,zinnjustin1980}. Here MMFT and RG represent the {\it modified mean field theory} proposed in this research paper and the {\it renormalization group}, respectively.\vspace{0.05in}}
\begin{ruledtabular}
\begin{tabular}{cccccc}\vspace{-0.05in}\\
Exponents&Experiment&RG ($n=1$)&Landau theory&MMFT ($k=0.47$)&Upper \& lower limits \vspace{0.05in} \\
\hline
\vspace{-0.05in}\\
$\alpha$&$0.11\sim0.12^{[a]}$&$0.110^{[c]}$&$0$&$0.113$&$1>\alpha>0$\\\vspace{-0.1in}\\
$\beta$&$0.28\sim0.38^{[b]}$&$0.325^{[c]}$&$\frac{1}{2}$&$0.34$&$\frac{1}{2}>\beta>\frac{1}{3}$\\\vspace{-0.1in}\\
$\gamma$&$1.20\sim1.47^{[b]}$&$1.241^{[c]}$&$1$&$1.47$&$1.5>\gamma>1$\\\vspace{-0.1in}\\
$\delta$&$4.6\sim4.9^{[a]}$&$4.82^{[a]}$&$3$&$4.88$&$5>\delta>3$\\\vspace{-0.1in}\\
\end{tabular}
\end{ruledtabular}
\end{table*}

At the critical point ($t=0$), we apply a weak electric field $E$ to the considered material, the exact differential of its Gibbs free energy will be $dF=-SdT+EdD=-SdT+EdP$; here $D$ represents the electric displacement and $D=\varepsilon_{0}E+P$, $\varepsilon_{0}$ is the electric permittivity of free space. $E$ is then written as
\begin{equation}
E=\left(\frac{\partial F}{\partial P}\right)_{T}=b(1-k)^{2}P^{3}\approx bP_{m}^{3+4k}.
\label{totaldifferential}
\end{equation}
Thus, the critical exponent $\delta$ can be defined as
\begin{equation}
\delta=3+4k.
\label{delta}
\end{equation}
If we let $k=0$, then $\delta$ will reduce to $\delta_{Landau}$. The lower limit of $\delta$ could then be regarded as 3; since $k<0.5$, its upper limit should be 5.

By taking the partial derivative of $E$ with respect to $P$, we get the following formula
\begin{equation}
\left(\frac{\partial E}{\partial P}\right)_{T}=\left(\frac{\partial^{2}F}{\partial P^{2}}\right)_{T}=a(1-k)+3b(1-k)^{2}P^{2}.
\label{gamma1}
\end{equation}
Then the corresponding electric susceptibility $\chi_{e}$ is given below
\begin{equation}
\chi_{e}=\frac{1}{\varepsilon_{0}}\frac{\partial P}{\partial E}=\frac{1}{\varepsilon_{0}}\frac{1}{a(1-k)+3b(1-k)^{2}P^{2}}.
\label{gamma2}
\end{equation}
When $t>0$, the order parameter $P$ will be zero. However, the polarization generated by the nematic phase will not be zero. In the vicinity of $T_{c}$, $k\rightarrow5$; thus the nematic phase has grown up, due to Le Chatelier's principle, and its volume is too large to respond to the applied weak electric field so that the remaining counteractive polarization does not contribute to $\chi_{e}$ and can be neglected. Then Eq. (\ref{gamma2}) will reduce to
\begin{equation}
\chi_{e}=\frac{1}{\varepsilon_{0}}\frac{1}{a(1-k)}.
\label{gamma3}
\end{equation}
When $t<0$, by using $P^{2}=-\frac{a}{b(1-k)}$, Eq. (\ref{gamma2}) can be modified as
\begin{equation}
\chi_{e}=\frac{1}{2\varepsilon_{0}}\frac{1}{(-a)(1-k)}.
\label{gamma4}
\end{equation}
Since $a\ll1$ in the vicinity of $T_{c}$, we can modify $a(1-k)$ by exploiting the method previously used. We let $a(1-k)=1+(1+k)\left[\frac{1-k}{1+k}a-\frac{1}{1+k}\right]$. By using the binomial series expansion, we have the following relationship $a(1-k)\approx a_{m}^{1+k}$; here $a_{m}=\frac{1-k}{1+k}a-\frac{1}{1+k}+1=\frac{1-k}{1+k}a+\frac{k}{1+k}$. Eq. (\ref{gamma3}) is then rewritten as
\begin{equation}
\chi_{e}=\frac{1}{\varepsilon_{0}}\frac{1}{a_{m}^{1+k}}. \ \ \ \mbox{$t>0$}
\label{gamma5}
\end{equation}
The critical exponent $\gamma$ can be defined as
\begin{equation}
\gamma=1+k^{+},
\label{gammaa}
\end{equation}
where $k^{+}\equiv k\rightarrow0.5^{+}$. Similarly, we can also derive the expression of the critical exponent $\gamma'$, which is given below
\begin{equation}
\gamma'=1+k^{-}.
\label{gammab}
\end{equation}
If $k=0$, both $\gamma$ and $\gamma'$ will reduce to $\gamma_{Landau}$ and $\gamma'_{Landau}$, respectively; thus, the lower limit of both $\gamma$ and $\gamma'$ is 1. If we let $k=0.5$, we can get their upper limit, which is equal to 1.5. In addition, $k^{+}>k^{-}$, thus $\gamma$ is slightly larger than $\gamma'$.

We now consider the asymptotic behavior of the specific heat $C$, which is defined as $C=-T\frac{\partial^{2}F}{\partial T^{2}}$, in the vicinity of $T_{c}$. We first rewrite $C$ as: $\frac{C}{T_{c}}=-\frac{T}{T_{c}}\frac{\partial^{2}F}{\partial T^{2}}=-t\frac{\partial^{2}F}{\partial T^{2}}-\frac{\partial^{2}F}{\partial T^{2}}=-\frac{aa_{0}}{T_{c}^{2}}\frac{\partial^{2}F}{\partial a^{2}}-\frac{a_{0}^{2}}{T_{c}^{2}}\frac{\partial^{2}F}{\partial a^{2}}$. If $k=0$, by using $F$ given in Eq. (\ref{modifiedlandau}), we can calculate $C$ as follows: $\frac{C}{T_{c}}=0$ when $t>0$ and $\frac{C}{T_{c}}=\frac{(-a)a_{0}}{2bT_{c}^{2}}+\frac{a_{0}^{2}}{2bT_{c}^{2}}$ when $t<0$. Clearly, in both cases, $C$ does not diverge and only has a discontinuity $\frac{a_{0}^{2}}{2bT_{c}^{2}}$ at $t=0$. Thus, the corresponding critical exponents are usually defined as $\alpha_{Landau}=\alpha'_{Landau}=0$.

If $k\neq0$, however, both $\alpha$ and $\alpha'$ will not be zero due to the embedded first order structural transformation. When $t>0$, the order parameter $P$ given in Eq. (\ref{modifiedlandau}) will be zero; however, the counteractive polarization generated by the nematic phase will not be zero. By replacing $P^{2}$ with $-\frac{a}{b(1-k)}$, we can rewrite the free energy based on the remaining counteractive polarization as follows
\begin{equation}
F=F_{0}-\frac{a^{2}}{b}\frac{k^{2}}{4(1-k)^{2}}=F_{0}-\frac{(ah)^{2}}{b},
\label{disorderfreeenergy2}
\end{equation}
where $h=\frac{k}{2(1-k)}$. Let $ah=a(1+k_{m})=1+(1+k_{m})\left[a-\frac{1}{1+k_{m}}\right]=1+(1+k_{m})\left[a-\frac{2(1-k)}{k}\right]$; near $T_{c}$, $a\ll1$ and $k\rightarrow1$ as explained previously, thus $\frac{2(1-k)}{k}\rightarrow0$ and $\left|a-\frac{2(1-k)}{k}\right|<1$. Similarly, using the binomial series expansion, we can get the following relationship: $ah\approx a_{m}^{1+k_{m}}$; here $k_{m}=h-1=\frac{3k-2}{2(1-k)}$ and $a_{m}=a-\frac{1}{1+k_{m}}+1=a+\frac{3k-2}{k}$. Then the above equation can be rewritten as
\begin{equation}
F\approx F_{0}-\frac{a^{2+2k_{m}}_{m}}{b}.
\label{disorderfreeenergy3}
\end{equation}
It can be seen that $\frac{\partial^{2}F}{\partial a^{2}}=\frac{\partial^{2}F}{\partial a_{m}^{2}}$; thus the specific heat $C$ can be calculated as
\begin{eqnarray}
C & = & -\frac{aa_{0}}{T_{c}}\frac{\partial^{2}F}{\partial a^{2}}-\frac{a_{0}^{2}}{T_{c}}\frac{\partial^{2}F}{\partial a^{2}}=-\frac{aa_{0}}{T_{c}}\frac{\partial^{2}F}{\partial a_{m}^{2}}-\frac{a_{0}^{2}}{T_{c}}\frac{\partial^{2}F}{\partial a_{m}^{2}} \nonumber \\
  & \approx & J_{1}a_{m}^{1+2k_{m}}+\left(a_{0}J_{1}+J_{1}J_{2}\right)a_{m}^{2k_{m}},
\label{specificheat1}
\end{eqnarray}
where $J_{1}=\frac{a_{0}(2+2k_{m})(1+2k_{m})}{bT_{c}}$ and $J_{2}=-\frac{3k-2}{k}$. Now we have two parameters describing the slow and the fast asymptotic behavior of $C$, respectively; the former is $1+2k_{m}$ and the latter $2k_{m}$. In practice, we usually measure the slow asymptotic behavior; therefore, we can define the critical exponent $\alpha$ as
\begin{equation}
\alpha=-(1+2k_{m})=\frac{1-2k^{+}}{1-k^{+}}.
\label{alphaa}
\end{equation}
When $t<0$, the corresponding free energy can be written as
\begin{equation}
F=F_{0}+F_{L}-\frac{(-a)^{2}}{b}\frac{k^{2}}{4(1-k)^{2}}.
\label{disorderfreeenergy4}
\end{equation}
In this case, $F_{L}$ does not affect the critical behavior (see $\alpha'_{Landau}=0$) and can be neglected. Thus, the critical exponent $\alpha'$ could be derived, in the same manner, as
\begin{equation}
\alpha'=\frac{1-2k^{-}}{1-k^{-}}.
\label{alphab}
\end{equation}
If $k=0$, the values of both $\alpha$ and $\alpha'$ will become a constant value of 1, which is their upper limit; if we let $k=0.5$, we will get their lower limit, which is equal to 0. We now try to determine which exponent, $\alpha$ or $\alpha'$, is larger. Subtracting $\alpha'$ from $\alpha$, we have the following expression
\begin{equation}
\alpha-\alpha'=\frac{k^{-}-k^{+}}{(1-k^{-})(1-k^{+})}.
\label{alphac}
\end{equation}
Since $k^{-}<k^{+}$, therefore, $\alpha'$ should be slightly larger than $\alpha$.

So far, the mathematical expressions of the critical exponents have been derived from the ferroelectric phase transition. But, by using the same method, they can also be derived from other continuous phase transitions. We now try to estimate the exact values of the exponents. In the vicinity of $T_{c}$, the value of $k$ of a perfect single crystal should be slightly less than 0.5, which is its theoretical limit. In practice, due to the existence of defects, $k$ varies from material to material near the critical point. For most magnetic materials with relative simple crystalline structures, $k$ should be close to 0.5. However, for some perovskite systems with complicated crystalline structures, $k$ may vary over a wide range of values. For instance, a Spain-Ukraine group recently investigated the evolution of the ferroelectric transition with Se doping in a ferroelectric family $\mathrm{Sn_{2}P_{2}(Se_{{\it x}}S_{1-{\it x}})_{6}}$ ($0\leq x\leq1$); for $x$ varying from 0.2 to 0.3, the exponent $\alpha$ they measured ranges from 0.14 to 0.34 \cite{vysochanskii2011}, which corresponds to the values of $k$ varying from 0.4624 to 0.3976. In this paper, without loss of generality, we let $k\approx0.47$; then the corresponding critical exponents can be calculated and the results are summarized in Table 1. Clearly, our results are in good agreement with the experimental data, which were mainly obtained in magnetic materials, and the results of renormalization group calculations.

It might be interesting to briefly discuss the Ginzburg criterion \cite{ginzburg1960,landau1980a}, in which the spatial correlation of the order parameter is used to quantitatively determine when the Landau theory is valid. The problem with this criterion is that, near $T_{c}$, the dynamic behavior of the correlation could be significantly influenced and modulated by the cooperative movement of HTSPs, which is not considered in Ginzburg's model. Therefore, the Ginzburg criterion may not provide a complete physical picture on why the Landau theory is not valid under certain circumstances.

Concluding remarks - the theoretical consideration and derivation presented in this paper have shown that there are two atomic movements involved in the ferroelectric phase transition. These two dynamics correspond to the vibration of crystalline lattice and the evolution of a partially ordered nematic phase formed by the cooperative behavior of high-temperature structure precursors, respectively; they are not independent of each other but rather constitute a hierarchical dynamic structure. The vibration represents the fast dynamics and the evolution corresponds to the slow dynamics, which is constrained by the vibration. The slow dynamics not only ``{\it imposes}" the thermodynamic limit on the occurrence of broken symmetry near the critical point but also alters the phase transition behavior, which are demonstrated by the fact that the cooperative behavior of HTSPs could certainly shift the asymptotic behavior of the corresponding physical quantities defined in Eqs. (\ref{criticalexponents1}). Now, it is perhaps safe to say that it is the behavior of the slow dynamics that makes the Landau theory deviate from experimental observations. The studies presented in this paper also rise a question whether it is necessary to draw a strict separating line between first-order and second-order phase transitions. According to the classical definition, there is no latent heat involved in second order phase transitions \cite{stanley1987}; within the framework of the theory developed in this paper, however, the slow dynamics always involves a weak first order structural transformation so that, contrary to the Landau theory, there should be latent heat involved in continuous phase transitions. Finally, we would like to emphasize that the conclusions given in this paper should apply to general continuous phase transitions though they are drawn from the ferroelectric phase transition.

\begin{center}
\textbf{\small ACKNOWLEDGMENT}
\end{center}
The research presented here was sponsored by the State University of New York at Buffalo. I am very grateful to Professor Yulian Vysochanskii of Uzhgorod National University of Ukraine for sharing his experimental studies with me.
\end{document}